\documentclass{aa}
\usepackage[varg]{txfonts}
\usepackage[T1]{fontenc}
\usepackage{graphicx}	% Including figure files
\usepackage{xcolor}	% Including figure files
\usepackage{amsmath}	% Advanced maths commands
\usepackage{amssymb}	% Extra maths symbols
\usepackage{caption}    % for managing figures better
\usepackage{subcaption} % for managing figures better
\usepackage{multicol}  
\usepackage{booktabs}  
\usepackage{longtable}
\usepackage{footmisc}
\usepackage{hyperref}
\usepackage{upgreek}
\usepackage{orcidlink}
\usepackage{placeins}
\bibpunct{(}{)}{;}{a}{}{,} % to follow the A&A style
\usepackage{soul} 
\usepackage{ulem} 
\usepackage{dblfloatfix}
\usepackage{float}
\usepackage{wrapfig}

\definecolor{encolor}{HTML}{b3003b}

\definecolor{nlcolor}{HTML}{5574d4}

\newcommand{\Msun}{{\rm M_{\sun}}}

\newcommand{\kpc}{{\rm kpc}}

\newcommand{\stream}{\rm s}
\newcommand{\cgm}{\rm cgm}

\newcommand{\muG}{\upmu\mathrm{G}}

\begin{document}
\title{Magnetising galaxies with cold inflows}

\author{Nicolas Ledos\orcid{0000-0001-9699-8941}\inst{1}
    \and Evangelia Ntormousi \orcid{0000-0002-4324-0034}\inst{2}
    \and Shinsuke Takasao \orcid{0000-0003-3882-3945}\inst{3}
    \and Kentaro Nagamine\orcid{0000-0001-7457-8487}\inst{3,4,5,6}}     

\offprints{N. Ledos, \email{nicolas.ledos@unimib.it}}

\institute{Dipartimento di Fisica "G. Occhialini”, Universit\`{a} degli Studi di Milano-Bicocca, Piazza della Scienza 3, 20126 Milano, Italy
  \and Scuola Normale Superiore di Pisa, Piazza dei Cavalieri 7, 56126 Pisa, Italy
  \and Theoretical Astrophysics, Department of Earth and Space Science, Graduate School of Science, Osaka University, 1-1 Machikaneyama, Toyonaka, Osaka 560-0043, Japan
    \and Theoretical Joint Research, Forefront Research Center, Graduate School of Science, Osaka University, 1-1 Machikaneyama, Toyonaka, Osaka 560-0043, Japan
  \and Kavli IPMU (WPI), UTIAS, The University of Tokyo, 5-1-5 Kashiwanoha, Kashiwa, Chiba 277-8583, Japan
  \and Department of Physics and Astronomy, University of Nevada, Las Vegas, 4505 S. Maryland Pkwy, Las Vegas, NV 89154-4002, USA
  }

\date{Received -- / Accepted --}

\abstract{High-redshift ($z\sim2-3$) galaxies accrete circumgalactic gas through cold streams. Recent high-resolution MHD simulations of these streams showed a significant amplification of the intergalactic magnetic field in the shear layer around them.} {In this work we estimate the magnetisation of high-redshift galaxies that would result purely due to the accretion of already magnetised gas from cold streams.} {We use the mass inflow rates and saturated magnetic field values from cold stream simulations as input to a simple analytic model that calculates the galactic magnetic field purely from mass accretion.} {Our model predicts average magnetic field strengths that exceed $\muG$ values at $z\sim 2-3$ for inflow rates above $0.1 \,\Msun \mathrm{yr^{-1}}$.  For high inflow rates, our model results are consistent with the recent detection of a strong magnetic field in a $z\gtrsim 2.6$ galaxies.} {Within the assumptions of our simple model, magnetised cold streams emerge as a viable mechanism for seeding a dynamically important galactic magnetic field.}

\maketitle

\section{Introduction}
Local spiral galaxies are strongly magnetised, with a magnetic energy density in equipartition with turbulent kinetic and cosmic ray energy densities \citep{Beck2019}. 
Galactic dynamo theory \citep[see ][for review]{Brandenburg2023} has been very successful in explaining cosmic growth and the observed large-scale coherence of the field, predicting growth rates of the order of $\rm 2~Gyr^{-1}$ for the large-scale field \citep{Beck1996}. 
However, recent observations suggest field strengths $>\muG$ at high redshifts as $z\sim 1$ \citep{Bernet2008, Mao2017},  $z\sim 2.6$ \citep{Geach2023} and up to $z\sim 5.6$ \citep{Chen2024}.
Strongly magnetised galaxies at a cosmic time of about $1\text{--}2\,\mathrm{Gyr}$ pose a critical challenge: 
either the seed fields for the galactic dynamo are stronger than previously thought, or there is an additional magnetic field amplification process at high redshift.

Numerical simulations of cosmic structure formation predict that massive galaxies are fed by filamentary cold accretion, penetrating through the hot circum-galactic medium (CGM) \citep{Fardal2001, Keres2005, Dekel2006, Dekel2009}. Recent 
high-resolution simulations of these cold streams have focused on their small-scale structure and hydrodynamical stability \citep{ Mandelker2019, bib_MHD_Berlok_2019b, Aung2019, Mandelker2020a, Ledos2024,Hong2024,Aung2024}. Using magnetohydrodynamic simulations with radiative cooling, \citet{Ledos2024} have revealed that an nG-strength field can be amplified up to $\muG$ values in the cold gas due to the tangling and stretching of the field lines inside the mixing layer between the stream and the CGM.

In light of this result and the observational evidence for high-$z$ galactic magnetisation, in this work we estimate the degree of magnetisation that can be obtained from magnetised gas inflow onto early galaxies. 
To this end, we construct a simple analytical model, quantifying the build-up of magnetic energy in the galaxy from magnetised cold inflows.

First, we present the origin of the magnetic field amplification from the simulations of \citet{Ledos2024} and from new 3D simulations in Section~\ref{sec:orig_mag}.
Then, we present the model in Section~\ref{sec:model_description} and discuss the results in Section~\ref{sec:results}.

\section{Magnetic field growth in cold streams}\label{sec:orig_mag}
Among recent high-resolution simulations of cold streams in the CGM with idealized geometry, \citet{Ledos2024} is so far the only one to cover the evolution of cold streams with magnetic field and radiative cooling over a wide range of parameters. 
We use their simulation results as input for our analytical model.
We briefly present the simulation suite below, along with new 3D simulations. For an extensive description, we refer the reader to \citet{Ledos2024}.

\subsection{The simulation setup} \label{sec:sim_description}

\begin{figure*}[h]
\centering
    \includegraphics[width=2.\columnwidth]{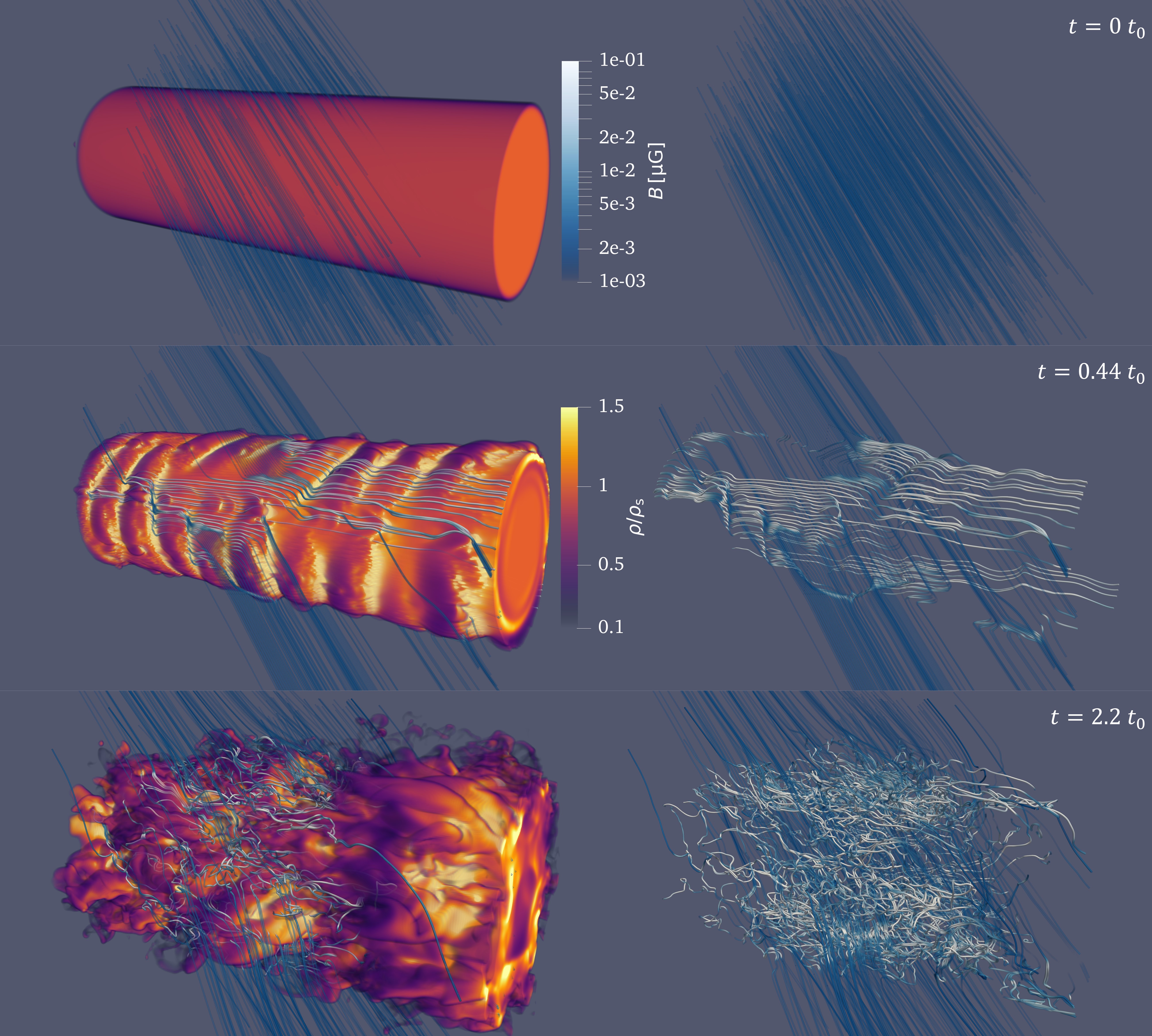}
    \caption{Time evolution of the stream and the magnetic field lines for a 3D simulation (with $n_{\stream}=0.01\, \mathrm{cm^{-3}}$). {\it Left:}The maps show volume rendering of the density field and the magnetic field lines. The lines are colored by magnetic field strength. {\it Right:} Same figure as the left one with only the magnetic field lines.}
    \label{fig:t_evol_3D}
\end{figure*}

The stream is positioned at the centre of a box defined by $L_{\mathrm{x}}\times L_{\mathrm{y}} \left( \times L_{\mathrm{z}}\right)= 64 R_{\stream} \times 32 R_{\stream} \left( \times 64 R_{\stream}\right)$ with $R_{\stream}$ representing the stream radius. The stream axis is oriented in the y-direction. 
The stream traverses the simulation box repeatedly due to periodic boundary conditions. Initially, the cold stream and the hot CGM gas are uniformly distributed in isobaric equilibrium, with a density contrast $\delta = \rho_{\stream} / \rho_{\cgm}$. This setup is consistent with previous idealised cold stream simulations \citep{Mandelker2016, Mandelker2019, bib_MHD_Berlok_2019b, Vossberg2019, Mandelker2020a}.

The initial conditions for the density and the magnetic field are shown in the top panels of Fig.~\ref{fig:t_evol_3D}. 
The physical properties are assumed to match those of the stream and CGM at the virial radius $R_{\mathrm{v}}=100\,\kpc$ of a typical massive halo of $10^{12} \, \mathrm{\Msun}$ at $z\sim 2$, consistent with cosmological simulations \citep{Goerdt2010}, empirical models \citep{Dekel2013,LyblobMandelker2020}, and observations \citep[e.g.,][]{Martin2012, Daddi2021, Emonts2023, Zhang2023}. This yields $n_{\stream} = 10^{-3}, 10^{-2}, 10^{-1} \, \mathrm{cm}^{-3}$, along with two density ratios $\delta \equiv \rho_{\stream} / \rho_{\cgm} = 30, 100$, a set of stream/CGM metallicities $\left(Z_{\stream}, Z_{\cgm}\right) = \left(10^{-1.5}, 10^{-1}\right) \, \mathrm{Z}_{\odot}$, and three different stream Mach numbers $\mathcal{M} = \mathcal{M}_{\cgm} = 0.5, 1, 2$. We fix the stream radius to  $R_{\stream} = 1\, \mathrm{kpc}$. In all cases, the magnetic field is initially uniform with a $\sim \mathrm{nG}$ strength, which yields a plasma beta (ratio of thermal pressure to magnetic pressure) $\beta = p / p_{\mathrm{mag}} = 10^5$. Magnetic fields with angle of $0^{\circ}$, $45^{\circ}$ and $90^{\circ}$ with respect to the stream are also considered.
Initial perturbations of the velocity field are introduced at the interface of the stream and the CGM.

Since magnetic fields can behave differently in 3D than in 2D, we ran two 3D simulations to validate our 2D results, extending the 2D stream into a 3D cylinder.
The 3D simulations are initialized with $n_{\mathrm{\stream}} = \left\{0.001,0.01\right\} \, \mathrm{cm}^{-3}$, $\delta = 30$, and $\mathcal{M} = 1$, employing a radiative cooling model from \citep{Ferland2017} without self-shielding 
or thermal conduction to reduce computational time\footnote{As demonstrated in \citet{Ledos2024}, thermal conduction has only a significant effect on magnetic field growth when $\mathcal{M} \leq 0.5$ and $n_{\mathrm{\stream}} \leq 0.001 \, \mathrm{cm}^{-3}$.}.

It should be noted that $\delta = 30$ represents a limit for the CGM properties, leading to relatively low temperature ($T_{\cgm}\sim 3\times 10^5 \, \mathrm{K}$) and sound speed ($c_{\cgm}\sim 120 \, \mathrm{km\, s^{-1}}$) and high density ($n_{\cgm}\sim 3\times 10^{-5} \text{--}3\times 10^{-3} \, \mathrm{cm}^{-3}$) \citep{Dekel2006,LyblobMandelker2020}. However, this value of $\delta$ does not significantly affect the magnetic field growth and the stream's evolution \citep{Ledos2024}, and has the advantage of reducing the computational cost of the simulations.

\subsection{Stream evolution}
Figure~\ref{fig:t_evol_3D} illustrates the evolution of the stream density and of the magnetic field lines in the 3D simulation with $n_{\mathrm{\stream}} = 0.01 \, \mathrm{cm}^{-3}$. The time is in units of the sound crossing time in the stream $t_0 = R_{\stream}/c_{\stream}$, with $c_{\stream}$ the sound speed in the stream. For the simulations in Fig.~\ref{fig:t_evol_3D}, $t_0\sim 45\, \mathrm{Myr}$, with a virial crossing time of $t_{\mathrm{v}}=R_{\mathrm{v}}/ v_{\stream,0} \sim 18\, t_0$.
As the stream progresses, the shear at the interface between the stream and the CGM triggers the Kelvin-Helmholtz instability (KHI). The growth of the KHI creates a mixing layer.
The fate of the stream is determined by the ratio of the cooling time in the mixing layer over the dynamical time,
\begin{equation} \label{eq:xi}
\xi =\frac{t_{\rm{cool}}}{t_{\rm{dyn}}},
\end{equation}
with $t_{\rm{cool}} = \left(T_{\rm{mix}} k_{\rm{b}}\right) /  \left[\left(\gamma -1 \right)n_{\rm{mix}}\Lambda_{\rm net,mix}\right]$, and $t_{\mathrm{dyn}} \sim R_{\stream} / v_{\stream,0}$, where $T_{\rm{mix}}$, $n_{\rm{mix}}$, and $\Lambda_{\rm net,mix}$ are the mixing layer gas temperature, number density and net cooling rate, estimated as in \citet{Begelman1990,Hillier2019}. The term $v_{\stream,0}$  is the stream velocity.
If cooling dominates, i.e., $\xi \lesssim 1$, then the CGM gas in the mixing layer condenses onto the stream, leading to a growth of the stream mass. 
This situation is illustrated in Fig.~\ref{fig:t_evol_3D}, where the simulations shown have $\xi \sim 6 \times 10^{-2}$. In cases where cooling is not dominant, \citet{Ledos2024} show that magnetic fields and thermal conduction can stabilise the stream against KHI up to $\xi \sim 8$. For higher $\xi$ values, the thermal conduction timescale becomes shorter than $t_{\rm{cool}}$ and $t_{\rm{dyn}}$, causing the stream to diffuse into the hot CGM, reducing its cold gas accretion rate.

\subsection{Magnetic field amplification mechanisms}
\citet{Ledos2024} showed that, in the situation described in the previous section, two mechanisms lead to magnetic field growth:
(1) the coherent stretching of the field lines due to the velocity shear at the interface between the stream and the CGM, (2) the tangling of field lines due to the growth of the KHI (including a incoherent stretching of field lines by turbulence/vortexes). We confirm that mechanisms (1) and (2) are dominant and well-resolved in our simulations by analysing the power spectra in Appendix ~\ref{app:power_spectra}. 
Additionally, the condensation of CGM gas onto the stream can lead to a compression mechanism that we discuss at the end of this section.

Figure~\ref{fig:t_evol_3D} shows the stretching and tangling of the magnetic field lines around the stream. The magnetic field is initially amplified by the stretching of the field lines at $t \sim 0.4 \mathrm{t_0} \sim 0.02 t_{\mathrm{v}}$. Following the growth of the KHI, the field lines tangle, and the magnetic field is mixed within the stream. This tangling has already developed well at the early time of $t \sim 2 \mathrm{t_0} \sim 0.1 t_{\mathrm{v}}$. Appendix~\ref{app:comp_2D} compares the results between the 2D and 3D simulations. We demonstrate that at later simulation times, the 3D simulations allow a more efficient mixing of the magnetic field in the stream compared to 2D simulations with the same flow parameters.
Some simulations in \citet{Ledos2024} have a magnetic field parallel to the stream, which results in a magnetic field growth only by tangling as there is no magnetic field component to be coherently stretched by shear at the interface of the CGM and the stream. We refer to this configuration as a no--stretching amplification case, which represents a lower limit for the magnetic field amplification.

The field amplification is approximated as a stretching flux tube \citep[see][for example]{Spruit2013}. The details of the model are derived and discussed in \citet[][Appendix D]{Ledos2024}. The magnetic field growth takes the form of
\begin{equation}\label{eq:Bmodel1}
    \frac{B_{\rm{th}}\left(t\right)}{B_0} = \frac{1}{2}\left[ \left(\beta \frac{R_{\stream}}{R_{\stream} + v_{\mathrm{c}}t}\right)^2 + 4\left(\beta+1\right)\right]^{1/2} - \frac{1}{2} \beta \frac{R_{\stream}}{R_{\stream} + v_{\mathrm{c}}t} ,
\end{equation}
where $B_{\rm{th}}$ is the theoretical magnetic field, and $\beta =10^5$ the plasma beta.
The term $R_{\stream}/\left(R_{\stream}+v_{\mathrm{c}}t\right)$ indicates the stretching of the field lines over time, with $v_{\mathrm{c}}$ 
the characteristic velocity of the mechanism responsible for the amplification.
In the case where field lines are stretched due to shear (mechanism 1), $v_{\mathrm{c}}=v_{\stream,0}$, i.e., the shear velocity between the stream and the CGM. In the case of the tangling of the field lines due to mixing (no--stretching case), $v_{\mathrm{c}}=v_{\mathrm{turb}}$ \citep[Eq. 34][]{Ledos2024}, where $v_{\mathrm{turb}}$ denotes the turbulent velocity in the mixing layer.
\\
\begin{figure}
\centering
    \includegraphics[width=1.\columnwidth]{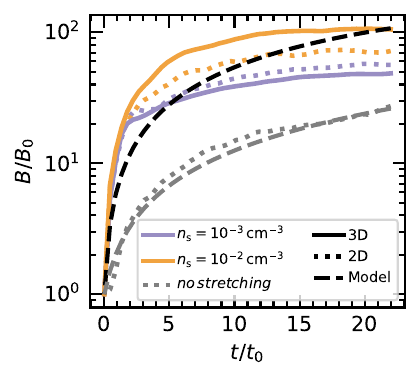}
    \caption{Time profiles of the magnetic field in the stream for 2D and 3D simulations in dotted and solid lines, respectively. Simulations are shown for $n_{\stream}= 10^{-2}$ and $10^{-3}\, \mathrm{cm^{-3}}$ in purple and orange, respectively. The dotted grey line stands for a 2D simulation with $n_{\stream}= 10^{-2}\, \mathrm{cm^{-3}}$ and with a magnetic field parallel to the stream (no amplification due to stretching). The dashed lines show the model results from Eq.~\ref{eq:Bmodel1} with the same Mach number as the simulations ($\mathcal{M} = 1$).}\label{fig:bvst}
\end{figure}

In Fig.~\ref{fig:bvst}, we present the time profile of the mean magnetic field in the stream for the 3D and 2D simulations, along with the model derived from Eq.~\ref{eq:Bmodel1}.
The magnetic field growth follows the same functional dependence on time in 2D and 3D. However, when cooling dominates ($n_{\stream}=10^{-2}\, \mathrm{cm^{-3}}$), the 3D simulation predicts a higher magnetic field strength compared to its 2D counterpart. This is because in 3D the mixing layer surrounds the stream, leading to a larger surface area for the stream to incorporate magnetic fields through the condensation of CGM gas. When condensation is subdominant  ($n_{\stream}=10^{-3}\, \mathrm{cm^{-3}}$), there is no significant difference of magnetic strength between the 2D and 3D simulations.
Given that the model (Eq.~\ref{eq:Bmodel1}) is independent of $n_{\stream}$, it reasonably reproduces the magnetic field growth in terms of shape and order of magnitude.
Fig.~\ref{fig:bvst} also depicts the magnetic field growth for a 2D simulation in the no--stretching case (only tangling and incoherent stretching, corresponding to a magnetic field initially parallel to the stream). By replacing the shear velocity $v_{\stream,0}$ in Eq.~\ref{eq:Bmodel1} with the turbulent velocity, $B_{\mathrm{th}}$ matches well the simulation result (compare the grey dotted and dashed lines). This highlights that in the absence of stretching, the primary mechanism for the field amplification is the tangling of the magnetic field lines due to turbulence.

Following these results, we consider that the magnetic field growth due to gas condensation is, in most cases, small compared to the effects of stretching and tangling. If compression was the dominant mechanism, the magnetic field growth would primarily depend on the efficiency of cooling in the mixing layer, i.e., the parameter $\xi$, and there would be almost no magnetic field amplification when the stream does not condense CGM gas ($\xi > 1$). However, (1) the magnetic field growth is both significant and independent of $\xi$ when the magnetic field is not parallel to the stream \citep[][figure 12]{Ledos2024}, and (2) when the magnetic field is parallel to the stream, the magnetic field growth can be well described by our model that accounts for the tangling of the field lines (see figure~\ref{fig:bvst}). The dependence on $\xi$ then arises from the turbulent velocity $v_{\mathrm{turb}}$. Based on these two arguments, we argue  that compression is 
%negligible (1) or 
subdominant %(2), 
and do not include it in our simple model.

\section{A simple Analytical Model}\label{sec:model_description} 
In this section, we present a simple analytical model to estimate the potential contribution of magnetised cold streams to the galactic magnetic field.
\subsection{Model summary}
\begin{figure*}
\centering
    \includegraphics[width=2.\columnwidth]{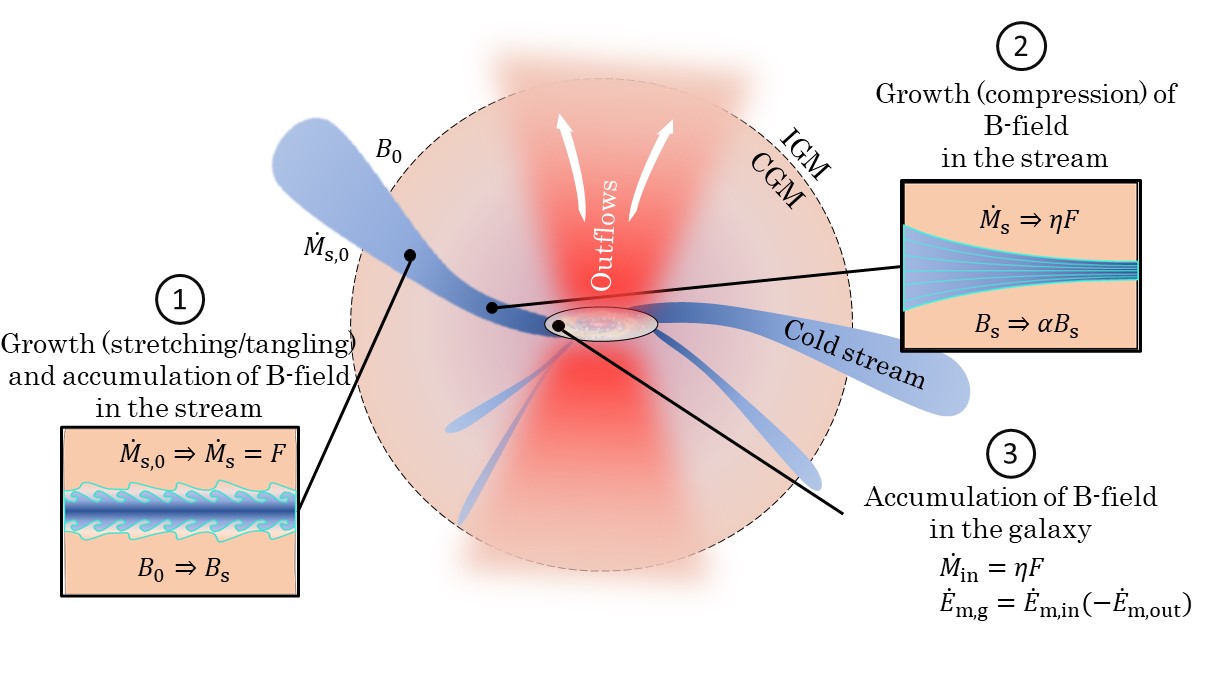}
    \caption{Illustrative sketch of the main processes of the model. The cold stream arrives initially from the intergalactic medium (IGM) with a mass flow rate $M_{\stream,0}$ and a magnetic field $B_0\sim \mathrm{nG}$. Step 1: As the cold stream enters the halo, a mixing layer forms at the interface of the CGM and the stream. Magnetic field lines are tangled and stretched in the mixing layer, resulting in magnetic field amplification. Simultaneously, CGM gas in the mixing layer condenses onto the stream, accumulating magnetic fields. From the interaction of the stream and the CGM, the cold mass flux and the magnetic field of the stream become $F$ and $B_{\stream}$. Step 2: Upon falling into the halo, the cold stream narrows and is squeezed by the dark matter potential. Consequently, magnetic fields in the stream are also compressed, further amplifying it by a factor $\alpha$. The stream is also expected to entrain CGM gas while falling through the halo potential, increasing its mass flux by an efficiency factor $\eta$. Step 3: Ultimately, if the stream reaches the galaxy, it continuously accretes cold gas at a rate $\dot{M}_{\mathrm{in}}$ with an associated magnetic energy inflow rate $\dot{E}_{\mathrm{m,in}}$ to the galaxy, elevating the galactic mean magnetic field. By default, we do not consider outflow, however, we introduce a simple outflow term in Appendix ~\ref{app:deriv_out}.}
    \label{fig:manga}
\end{figure*}

We illustrate the main idea of the model in Fig.~\ref{fig:manga}.
The model can be decomposed into three physical steps:
\begin{itemize}
    \item Step 1: Growth and accumulation of the magnetic field in the stream. \quad
    Once the cold stream enters the hot halo, the velocity shear at the interface of the cold stream and the hot CGM triggers the KHI. From the growth of the KHI, a mixing layer forms at the interface. The magnetic field lines are tangled and stretched by shear and KHI-generated turbulence, leading to an amplification of the magnetic field in the mixing layer.  Concurrently, if the cooling time in the mixing layer is shorter than the shearing time, the CGM gas condenses onto the stream. The condensation incorporates the magnetic field from the mixing layer into the cold stream. This process occurs relatively fast, with $t\lesssim8t_0 \sim 0.4 t_{\mathrm{v}}$.    
    As a consequence of the interaction between the stream and the CGM, the amplified magnetic field $B_{\stream}$ within the stream and the stream's cold mass flux $\dot{M}_{\stream}$ are expressed as follows:
    \begin{equation}\label{eq:Bs}
       B_{\stream} = B_{\stream}\left(\mathbf{X}\right),
    \end{equation}
    \begin{equation}\label{eq:F}
        \dfrac{\mathrm{d}M_{\mathrm{s}}}{\mathrm{d}t} = F\left(\mathbf{X}\right).
    \end{equation}
    To keep generality, one may use any functions for $B_{\stream}\left(\mathbf{X}\right)$ and $F\left(\mathbf{X}\right)$ with  $\mathbf{X}$ being a vector of any relevant inputs variables. The specific form of these functions is defined in Sect.~\ref{sec:sim2model} within the context of our simulations.   
    \vspace{2mm}
    \item Step 2: Growth of the magnetic field in the stream from compression. \quad 
    The cold stream falling towards the central galaxy undergoes compression due to the gravitational potential of the dark matter halo. This compression is expected to amplify the magnetic field within the stream further.  
    Additionally, as the stream grows due to the condensation of CGM gas, its velocity increases, resulting in a higher rate of cold gas accretion onto the galaxy.
    Our suite of simulations, presented in Sect.~\ref{sec:orig_mag},
    does not account for this process, but we include it in the analytical model by considering the results from other works.
    Recent studies from \citet{Hong2024,Aung2024} show that the potential can indeed enhance the cold gas accretion rate by a factor of $\lesssim 3.5$.    
    Because of the halo potential's influence, we define the cold mass flux $\dot{M}_{\mathrm{in}}$ reaching the galaxy as
    \begin{equation}\label{eq:Min}
        \dfrac{\mathrm{d}M_{\mathrm{in}}}{\mathrm{d}t} = \eta F\left(\mathbf{X}\right),
    \end{equation}
    and its associated mean magnetic field $B_{\mathrm{in}}$ as
    \begin{equation}\label{eq:Bin}
       B_{\mathrm{in}} = \alpha B_{\stream}\left(\mathbf{X}\right)
    \end{equation}
    where $\eta\in [0,\infty]$ represents an arbitrary efficiency rate for the cold inflow due to the halo potential, and $\alpha$ is the compression coefficient that accounts for the magnetic field amplification resulting from compression.    
    \vspace{2mm}
    \item Step 3: Accumulation of magnetic energy in the galaxy. \quad 
    We assume that the galaxy is fed by magnetised cold inflows with accretion rate $\dot{M}_{\mathrm{in}}$ and magnetic field $B_{\mathrm{in}}$.  Over time, this inflow accumulates magnetic energy within the galaxy, thereby increasing the mean galactic magnetic field $B_{\mathrm{g}}$. 
    Clearly, galaxies can also lose magnetic flux through outflows, an effect not included in the model.
    Since outflow properties depend on the galactic properties in a non-trivial way, including them would require extensions to the model that are beyond the scope of this work. However, in Appendix~\ref{app:deriv_out}
    we consider the simple case 
    of a continuous outflow term.
    More detailed estimates of the effect of outflows on the galaxy's magnetisation will be the subject of future work.
\end{itemize}

\subsection{Derivation}
A general definition of the evolution of the magnetic energy in the galaxy can be written as,
\begin{equation}\label{eq:emg_gen2}
    \dfrac{\mathrm{d} E_{\mathrm{m,g}}}{\mathrm{d} t} = \dfrac{\mathrm{d} E_{\mathrm{m,in}}}{\mathrm{d} t} + \dfrac{\mathrm{d} E_{\mathrm{m,amp}}}{\mathrm{d} t} - \dfrac{\mathrm{d} E_{\mathrm{m,diff}}}{\mathrm{d} t} - \dfrac{\mathrm{d} E_{\mathrm{m,out}}}{\mathrm{d} t},
\end{equation}
where the terms $\dot{E}_{\mathrm{m,in}}$, $\dot{E}_{\mathrm{m,amp}}$, $\dot{E}_{\mathrm{m,diff}}$, and $\dot{E}_{\mathrm{m,out}}$ stand for the variation of galactic magnetic energy by inflows, amplification inside the galaxy, diffusion processes, and removal from outflows, respectively.
For simplicity, in this paper, we focus only on the inflow component $E_{\mathrm{m,in}}$, leaving all other terms equal to zero. We provide the derivation of the model with a simple outflow component $E_{\mathrm{m,out}}$ in Appendix~\ref{app:deriv_out}. 
We can then rewrite $\dot{E}_{\mathrm{m,g}}$ as
\begin{equation}
    \dfrac{\mathrm{d} E_{\mathrm{m,g}}}{\mathrm{d} t} = \dfrac{\mathrm{d} E_{\mathrm{m,in}}}{\mathrm{d} t} = \dfrac{\mathrm{d} E_{\mathrm{m,in}}}{\mathrm{d} M_{\mathrm{in}}}\dfrac{\mathrm{d} M_{\mathrm{in}}}{\mathrm{d} t}.
\end{equation}
Knowing that,
\begin{equation}\label{eq:emin}
    E_{\mathrm{m,in}}= \frac{B_{\mathrm{in}}^2}{2}  V_{\mathrm{in}} = \frac{B_{\mathrm{in}}^2}{2} \frac{M_{\mathrm{in}}}{\rho_{\mathrm{in}}},
\end{equation}
where $V_{\mathrm{in}}$, $M_{\mathrm{in}}$ and $\rho_{\mathrm{in}}$ are the volume, mass, and density of the gas inflow reaching the galaxy, respectively.
We end up with a general formulation as,
\begin{equation}\label{eq:dEmgdt2}
    \dfrac{\mathrm{d} E_{\rm{m,g}}}{\mathrm{d} t} = \frac{B_{\mathrm{in}}^2}{2 \rho_{\mathrm{in}}} \dfrac{\mathrm{d} M_{\mathrm{in}}}{\mathrm{d} t} = C_{\mathrm{in}}.
\end{equation}
Assuming that $C_{\mathrm{in}}$ is invariant in time yields to,
\begin{equation}\label{eq:Emg}
    E_{\mathrm{m,g}}(t) = C_{\mathrm{in}} t +  E_{\mathrm{m,g,0}},
\end{equation}
For $C_{\mathrm{in}}$, using Eqs.~\ref{eq:Min}--\ref{eq:Bin} for $B_{\mathrm{in}}$ and $\dot{M}_{\mathrm{in}}$ gives
\begin{equation}\label{eq:cin}
    C_{\mathrm{in}}= \frac{\alpha B_{\mathrm{s}}^2}{2 \rho_{\mathrm{s}}} \eta F\left(\mathbf{X}\right)
\end{equation}
where we use the compression coefficient such that $\rho_{\mathrm{in}}\equiv \alpha \rho_{\stream}$.

Finally, the mean magnetic field in the galactic volume $V_{\rm{g}}$ can be recovered as,
\begin{equation}\label{eq:Bg2}
    B_{\rm{g}}(t) = \left(2\frac{ E_{\mathrm{m,g}}(t)}{V_{\rm{g}}} \right)^{1/2}.
\end{equation}
From Eq.~\ref{eq:Bg2}, the  model predicts a continuous growth of the galactic magnetic field.

The above derivation is kept general such that any estimation of the cold mass inflow $\dot{M}_{\mathrm{in}}$, and the inflow magnetic field $B_{\mathrm{in}}$ can be used and plugged in either in Eq.~\ref{eq:dEmgdt2} or Eq.~\ref{eq:Bg2}.

\subsection{Step 1: Input values from simulations}
\label{sec:sim2model}
We now outline how we utilize the simulations to determine values for the cold stream mass flux $F$ and its magnetic field $B_{\stream}$ resulting from the interaction with the CGM (Step 1). To determine the cold mass flux subsequent to the interaction with the CGM, we employ a fitting function based on the cold mass flux rate obtained from the 2D simulations suite \citep{Ledos2024}. This function is expressed as follows:

\begin{equation}\label{eq:fitF}
    \frac{F\left(\mathbf{X}\right)}{\dot{M}_{\stream,0}} = 1 - a\,\xi^b, 
\end{equation}
where $a\sim\left\{0.03,0.11,0.05\right\}$ and $b\sim\left\{0.95,0.24,0.12\right\}$, for $\mathcal{M}=\left\{0.5,1,2\right\}$, respectively. Here, $\mathbf{X}=\left(\mathcal{M},\xi \right)$.  It is important to note that the fits are performed based on the mass flux at $t = t_{\mathrm{end}}$, which therefore represents a lower limit scenario for the cold mass survival rate.

\begin{figure}
    \centering
    \includegraphics[width=1.0\columnwidth]{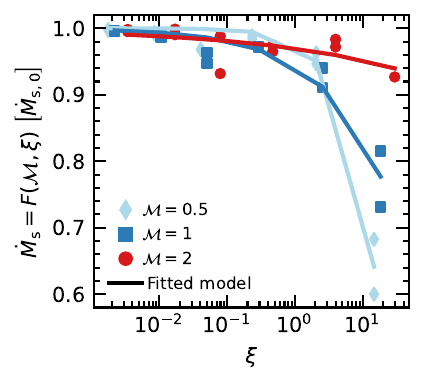}
    \caption{Cold mass flux rate normalised by the initial cold mass flux of the simulations. Each point represents one simulation with a fixed $\xi$, $\mathcal{M}$, and initial magnetic angle $\phi$ with respect to the stream axis. Red dots, blue squares, and cyan diamonds refer to 2D simulations with $\mathcal{M}= 2,1,0.5$, respectively. Using the same colour legend, the fitted model is represented by coloured curves. For a fixed $\xi$ and $\mathcal{M}$, the two points represent simulations with an initial magnetic field angle of $90^{\circ}$ and $45^{\circ}$ with the highest values are associated with the $90^{\circ}$ angle in all cases.}
    \label{fig:mdot}
\end{figure}

The comparison of the fitted model and the simulations is shown in Fig.~\ref{fig:mdot}.
As described in Sect.~\ref{sec:sim_description}, the cold stream mass flux is expected to decrease with increasing $\xi$. Moreover,  for $\xi\gtrsim 1$, higher Mach numbers correspond to lower KHI growth rates, resulting in higher values of $F$ in the simulations. The fitted model from Eq.~\ref{eq:fitF} effectively predicts the simulation outcomes, demonstrating its suitability as a definition for $F$.

For $B_{\stream}$, we assume the magnetic field to be at the mean saturated value following the initial rapid growth, $B_{\stream}= \left\langle B_{\mathrm{th}}\right\rangle$, with the average done over the full simulation time $t=22t_0$ and with $B_{\mathrm{th}}$ expressed in Eq.~\ref{eq:Bmodel1} for $\xi<8$.
For $\xi>8$, to account for the special case where the stream is diffused inside the CGM due to thermal conduction, we use a linear function dependent on Mach numbers fitted from values of \citet{Ledos2024}. This gives,
\begin{equation}
B_{\stream} = \left\{ \begin{array}{ll}
\left\langle B_{\mathrm{th}}\right\rangle & \text{if} \, \, \xi \leq 8,\\
B_0 \left(34\mathcal{M}-16\right) & \text{if} \, \, \xi >8.
\end{array}\right.
\label{eq:Bs_default}
\end{equation}
For $\xi > 8$, anisotropic thermal conduction is expected to diffuse both the stream and the perturbations, thereby suppressing the KHI. Thermal conduction occurs along the magnetic field lines, which are stretched at the stream's interface due to the velocity difference. Faster streams result in more stretching and less efficient thermal conduction \citep[see][section~4.1.2]{Ledos2024}. Consequently, the stream’s diffusion continues over time only for relatively slow streams ($\mathcal{M} \leq 0.5$). For $\mathcal{M} > 0.5$, the stream's diffusion becomes rapidly inefficient, so the magnetic field, initially amplified by stretching, remains in the cold phase, leading to a small amplification of the mean magnetic field within the stream.

As discussed in Sect.~\ref{sec:orig_mag}, we also consider a no--stretching case for the magnetic field amplification. In that case, instead of using Eq.~\ref{eq:Bmodel1}, we want to account for the impact of thermal conduction and use power law fit from \citet{Ledos2024}, giving,
\begin{equation}
B_{\stream} \sim 0.13\,\xi^{-0.1}.
\label{eq:Bs_nostretching}
\end{equation}
Note that $B_{\stream}$ is based on the 2D simulations results \citep[see ][fig.~12]{Ledos2024}. As discussed in Sect.~\ref{sec:orig_mag}, this means that our $B_{\stream}$ may be an underestimation compared to our 3D results when cooling is dominant ($\xi<1$).

\section{Results and discussion}\label{sec:results}
We first apply our model to a general case of an accreting galaxy and then consider the particular case of the galaxy studied by \citet{Geach2023}.
\subsection{Growth of the galactic magnetic field}
We consider a scenario without outflows (Eq.~\ref{eq:Emg}) where the impact of the halo potential (Step 2) is neglected, i.e., $\eta=\alpha=1$.
For the galaxy volume $V_{\mathrm{g}}$, we first consider its halo mass to be on the order of $10^{12} \, \Msun$ to ensure that its CGM is hot and that it is fed by cold streams between $z\sim 3$ and $z\sim 2$ \citep{Dekel2006,Aung2024}. This redshift range corresponds to $\sim 1.15 \,\mathrm{Gyr}$, for which we assume for simplicity that the cold accretion happens for about 1 Gyr.
We use the empirical relation of \citet{Girelli2020} for the halo to stellar mass ratio, and the empirical relation of \citet{Yang2021} for the stellar mass to galactic radius conversion\footnote{
\citet{Girelli2020}'s empirical estimation of the stellar-to-halo mass relation is for star-forming galaxies up to redshift $z\sim 4$ from the COSMOS catalogue \citep{Scoville2007} and the {\tiny{DUSTGRAIN}}--{\it pathfinder} simulations \citep{Giocoli2018}. The empirical relation from \citet{Yang2021} is for massive star-forming galaxies up to $z\sim 2.75$ and uses the ASTRODEEP catalogue.} For our considered halo and redshift, it gives us a stellar mass of $M_* \sim 2\times 10^{11} \, \Msun$, and $R_{\mathrm{g}}\sim 3\, \kpc$. The galaxy's disc height is fixed at $d_{\mathrm{g}} = 0.1\, R_{\mathrm{g}}$,
%10\% \, R_{\mathrm{g}}$, 
as suggested by observations of high-redshift galaxies from the {\it Hubble} Space Telescope (HST) catalogue \citep[][figure 19]{Elmegreen2017}. Notably, their values of $d_{\mathrm{g}} / R_{\mathrm{g}}$ range from $\sim 2\%$ to $\sim 25\%$ at $z=2.5\text{--}2$. The effects of an extended height and radius are discussed in Section~\ref{sec:discussion}.

By fixing $\delta=30$ and $\mathcal{M}=1$, $F$ and $B_{\stream}$ become solely functions of $R_{\stream}$ and $n_{\stream}$, the cold stream radius and number density at the virial radius of the halo. Fixing $R_{\stream}$, the galactic magnetic field can then be plotted in function of $\dot{M}_{\stream,0}$ by varying only $n_{\stream}$.
\\
\begin{figure}
    \centering
    \includegraphics[width=1.0\columnwidth]{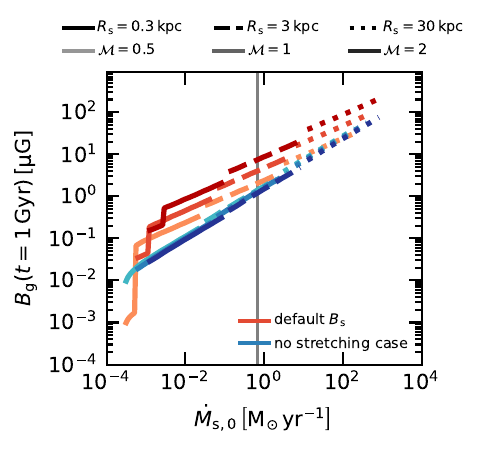}
    \caption{Mean galactic magnetic field from our model as a function of the stream cold mass flux entering the halo $\dot{M}_{\stream,0}$. The cold mass flux is obtained by varying its number density $n_{\stream}\in \left[0.001,0.1\right] \, \mathrm{cm^{-3}}$ for different stream radii $R_{\stream}=(0.3,3,30)\,\kpc$ (plain, dashed and dotted lines) and for different Mach number $\mathcal{M}=0.5,1,2$ (light to dark colours). The red and blue colour maps show results for the default case $B_{\stream}$ and the no--stretching case (see Sec~\ref{sec:sim2model}), respectively. The vertical grey line shows the maximum cold mass flux simulated in the simulations suite.}
    \label{fig:results1}
\end{figure}
Fig.~\ref{fig:results1} illustrates the galactic magnetic field strength after 1~Gyr of accretion by one cold stream of radius $R_{\stream}$ as a function of the cold stream mass flux. The results are plotted for three 
values of the stream radius and the two values of $B_{\stream}$ growth. 

From the fiducial model (default $B_{\stream}$), even a small cold stream of radius $R_{\stream}=0.3\, \kpc$ yields $\muG$ galactic magnetic field values for $\dot{M}_{\stream,0}\gtrsim 0.1 \, \Msun \mathrm{yr^{-1}}$. A lower $\dot{M}_{\stream,0}$ corresponds to a lower stream number density $n_{\stream}$ (higher $\xi$), resulting in a shorter thermal conduction timescale and a longer cooling time\footnote{To note, lowering the radius, the number density and the stream velocity can also lower $\dot{M}_{\stream,0}$ and increase the parameter $\xi$. For a detailed quantitative description of the timescale comparison depending on the stream parameters, refer to \citet[][section 2.5]{Ledos2024}.}. Consequently, for narrow streams with $R_{\stream}=0.3\, \kpc$, thermal conduction only diffuses the stream at relatively low accretion rates per stream of $\dot{M}_{\stream,0}<10^{-2} \, \Msun \mathrm{yr^{-1}}$. This stream diffusion inhibits the magnetisation of the galaxy only below this low accretion rate threshold. In the no--stretching case, the magnetic field amplification occurs for $R_{\stream}=3\, \kpc$ with $\dot{M}_{\stream,0}\gtrsim 1 \, \Msun \mathrm{yr^{-1}}$.

Constraining both the minimum size and the number of cold streams that feed massive galaxies remains a significant challenge, primarily due to the current resolution limits for cold gas in the CGM of cosmological simulations \citep[e.g.,][]{Bennett2020}. Given the connectivity between galaxies in cosmological simulations, it is expected that massive star-forming galaxies are fed by approximately two or more streams \citep{Espinosa2023}, with relatively large radii when entering the halo ($R_{\stream}\gtrsim 10\, \kpc$). The streams become narrower as they approach the galaxy due to the pressure gradient in the CGM and gravitational focusing caused by the halo potential. Such cold stream accretion picture is commonly predicted in simulations and empirical models \citep[e.g.,][]{Dekel2009,Rosdahl2012,Mandelker2018,LyblobMandelker2020,Lu2024}. With at least two large cold streams accreting onto a galaxy for about $1 \, \mathrm{Gyr}$, our model naturally predicts $B_{\mathrm{g}}\gtrsim 10\, \muG$ for both cases of $B_{\stream}$.

\subsection[temp]{Applying the model to the galaxy of \citet{Geach2018}}
We now apply our model to the observed lensed star-forming galaxy \citep{Geach2018} at redshift $z\sim 2.6$ whose magnetic field is estimated at $\sim 500\, \muG$ or less \citep{Geach2023}. 
The galaxy's dynamical mass is $M_{\mathrm{dyn}}\sim 8.1\times 10^{10}\, \Msun$, with a radius of about $2.5\, \kpc$ and a star-formation rate (SFR) of the order of $\sim 1000\, \Msun\mathrm{yr^{-1}}$. 
We assume that the galaxy's host dark matter halo follows an NFW profile. Then, assuming a polytropic equation of state $P_{\cgm}\propto \rho_{\cgm}^{\gamma'}$, the radial density profile follows \citep[][equation 19]{Komatsu2001}:
\begin{equation}
    y\left(r\right) =\frac{\rho_{\cgm}\left(r\right)}{\rho_{\cgm,0}} = \left[1 +\frac{3}{\sigma_0}\frac{\gamma'}{\gamma'-1}\frac{c}{m\left(c\right)}\left(\frac{\log{\left(x+1\right)}}{x}-1\right) \right]^{\frac{1}{\gamma'-1}},
\end{equation}
where $r$ is the radius, $c$ is the concentration parameter of the halo, $x\equiv rc/R_{\mathrm{v}}$, $m\left(c\right)=\log{\left(c+1\right)} + c/\left(c+1\right)$, and $\rho_{\cgm,0}=\rho_{\cgm}\left(r=0\right)$.
The constant $\sigma_0$ and the polytropic index $\gamma'$ are empirical functions\footnote{For $c=10$ and 1, this leads to $\sigma_0\sim 3.28, \gamma'\sim 1.19$ and $\sigma_0\sim 1.55, \gamma'\sim 1.10$, respectively \citep[][equations 25--26]{Komatsu2001}.} of $c$. Further assuming pressure equilibrium between the stream and the hot CGM gas $P_{\stream}=P_{\cgm}$ \citep{Aung2024}, we can define the compression parameter $\alpha$ as the ratio of the stream density between $0.1R_{\mathrm{v}}$ and $R_{\mathrm{v}}$:
\begin{equation}
    \alpha = \frac{\rho_{\stream}\left(0.1R_{\mathrm{v}}\right)}{\rho_{\stream}\left(R_{\mathrm{v}}\right)} = \left[\frac{y\left(0.1R_{\mathrm{v}}\right)}{y\left(R_{\mathrm{v}}\right)}\right]^{\gamma'}\frac{T_{\stream}\left(R_{\mathrm{v}}\right)}{T_{\stream}\left(0.1R_{\mathrm{v}}\right)},
\end{equation}
where we assume $T_{\stream}\left(R_{\mathrm{v}}\right) / T_{\stream}\left(0.1R_{\mathrm{v}}\right) \sim 1$\footnote{In equilibrium state, the cold stream's temperature is defined by the balance between radiative cooling and heating, leading to $T_{\stream}\sim 10^4 \, \mathrm{K}$. However, as the stream becomes denser inside the halo, $T_{\stream}$ can slightly decrease, leading to $T_{\stream}\left(R_{\mathrm{v}}\right) / T_{\stream}\left(0.1R_{\mathrm{v}}\right) \sim 1\text{--}2$ \citep{Aung2024}. Such an additional factor would slightly increase the parameter $\alpha$.}.

We define two scenarios for our model: a lower limit scenario and a fiducial scenario. For the lower limit scenario, the mass accretion is assumed at its minimum, such that $\dot{M}_{\stream,0}= \text{SFR} = 1000\, \Msun\mathrm{yr^{-1}}$. For the model parameters, the accretion efficiency $\eta$ and the concentration parameter $c$ are both set to unity leading to $\alpha\sim 22$. The no--stretching case is assumed for $B_{\stream}$. To note, $c=1$ represents only a numerical lower limit of the model and is a factor $2$--$4$ lower than estimation with $\Lambda$-CDM \citep[e.g., ][]{Prada2012,Dutton2014,Ludlow2016}.

For the fiducial scenario, we adopt a ratio of mass accretion rate $\dot{M}_{\mathrm{in}}$ over the star formation rate, $\epsilon = \dot{M}_{\mathrm{in}} / \text{SFR} \sim 0.2$. This ratio accounts for the fact that the cold gas accreted by the galaxy is not instantaneously transformed into stars, but instead contributes to the gas reservoir of the galaxy, sustaining star formation over longer timescales. The value of $0.2$ is motivated by those derived from observations by \citet[][$\epsilon \sim 0.12$]{Daddi2021}, \citet[][$\epsilon \sim 0.56$]{Emonts2023}, and \citet[][$\epsilon \sim 0.12$]{Zhang2023}. It should be noted that this ratio is not well constrained and can vary significantly, for example, during a galaxy's sudden starburst phase, where $\epsilon$ can reach $\sim 12$ for $\text{SFR} \sim 1200 \, \Msun\mathrm{yr^{-1}}$ \citep{Fu2021}. Further quantification of the ratio $\epsilon$ is left for future work, as part of the development of a less simplistic model. The resulting total accretion rate onto the galaxy is $\dot{M}_{\mathrm{in}} \sim \text{SFR} \, \epsilon^{-1} = 5000\, \Msun\mathrm{yr^{-1}}$. Following \citet{Aung2024}, we consider an efficiency parameter $\eta=3$, yielding $\dot{M}_{\stream,0}= \dot{M}_{\mathrm{in}} \eta^{-1} = 5000\, \eta^{-1} \, \Msun\mathrm{yr^{-1}}$ at the virial radius. We assume a concentration parameter $c=10$, which leads to a compression coefficient of $\alpha \sim 450$.

In both scenarios, we assume that the galaxy accretes through three cold streams with a number density at the virial radius of $n_{\stream}=0.1\, \mathrm{cm^{-3}}$\footnote{We fixed the number density for the stream at a relatively high value compared to predictions from empirical models \citep{LyblobMandelker2020,Aung2024} in order to avoid artificially large streams radii or numbers.}. This results in stream radii of $19 \, \kpc$ and $35 \, \kpc$ for the lower limit and fiducial scenarios, respectively. We also apply our simple model extension with outflow presented in Appendix~\ref{app:deriv_out} using a $\text{SFR} = 1000\, \Msun\mathrm{yr^{-1}}$.

\begin{figure}
    \centering
    \includegraphics[width=1.0\columnwidth]{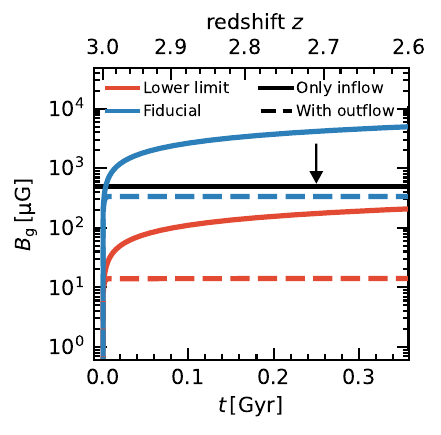}
    \caption{Time evolution of the mean galactic magnetic field of our model. The corresponding redshift is also given at the top axis assuming that the accretion starts at $z=3$. Blue and red curves represent the fiducial and lower-limit scenarios. Solid and dotted lines are models without and with outflows. The black line with the arrow shows the $500\, \muG$ or less target from \citet{Geach2023}.}
    \label{fig:results2}
\end{figure}

Figure~\ref{fig:results2} illustrates the time evolution of the galactic magnetic field $B_{\mathrm{g}}$ assuming that the accretions start at redshift $z=3$. In the lower limit scenario, owing to the extreme assumptions, the galactic magnetic field is raised to $\sim 250\,\muG$ values by $z=2.6$. With the reasonable assumptions of the fiducial model, the galaxy reaches $500\,\muG$ rapidly at around $t\sim 10 \, \mathrm{Myr}$. 
If we consider the expulsion of galactic magnetic energy by a constant outflow (see Appendix~\ref{app:deriv_out}), the initial growth is the same as before, however, the galactic magnetic field reaches a plateau. This behaviour illustrates the necessity to embed this inflow model into a more sophisticated framework that includes the complex physics of galactic outflows.

Within the limit of the model assumptions, cold streams emerge as plausible candidates for rapidly magnetising galaxies, potentially elucidating the puzzlingly large magnetic field estimates by \citet{Geach2023}.

\subsection{Discussion}\label{sec:discussion}
Numerous simplifications are assumed in our model, and we discuss here the potential impact of the missing mechanisms.
Firstly, the model assumes that the cold streams directly reach the galaxy. However, cosmological simulations suggest that cold streams have an impact parameter and circularize into an extended ring around the galaxy \citep{Danovich2015,Stewart2017}. Such an extended ring would likely delay the buildup of magnetic energy in the galaxy.
Additionally, upon reaching the galaxy or its outer ring, the cold streams are more likely to fragment, as suggested by observations of H$\alpha$ clumps in galaxies at redshift $z\sim 1.5\text{--}2$ \citep{Fisher2017} and simulations \citep{Ceverino2016,Mandelker2018,Aung2019}. These studies associate these clumps with the creation of turbulent galactic disks \citep{Ginzburg2022,Forbes2023}. Such fragmentation and turbulence could lead to further amplification of the magnetic field through mechanisms similar to our  Step 1 and Step 2, in addition to galactic dynamo processes. 
Additionally, we assume that cold accretion remains constant across redshifts. Theoretical and empirical models suggest a $(1+z)^{2.5}$ dependence of the accretion rate for $z \gtrsim 2$ \citep{Neistein2008,Fakhouri2010,Dekel2013}. While this would introduce only a factor of $\sim 1.2$ to the final $B_{\mathrm{g}}$ within the redshift range considered in section~\ref{sec:results}, such a redshift dependency might be necessary to accurately trace magnetic field evolution over a longer timescale. Extending the model to include this aspect is left for future work.

Our model only accounts for inflows. We have estimated the possible effect of an outflow from supernovae feedback only in a simplified and limited case. Considering a wider framework of galaxy evolution where supernova outflows arise self-consistently is an obvious next step for our model. Active galactic nuclei (AGN) feedback can also remove magnetic energy from the galaxy, dispersing it into the CGM. Incorporating an AGN outflow term into Eq.~\ref{eq:dEmgdt2} is a focus for future work. However, in the presence of outflows, it would be necessary to also consider re-accretion through the galactic fountain \citep[e.g.,][]{Putman2012} within the inflow term. Therefore, while we do not account for AGN feedback, not accounting for re-accretion might lead to an overall underestimation of the galactic magnetic field growth by our model.

Our model also does not explicitly solve any diffusion processes. The simulation suite from \citet{Ledos2024} explicitly solves thermal conduction, and its effects are implicitly considered in our definition of the cold gas flux $F$ after interaction with the CGM ( Step 1, Eq.~\ref{eq:fitF}). As the stream becomes denser throughout the halo, the thermal conduction timescale increases towards the centre of the halo. Consequently, its impact should be negligible for Step 2 and Step 3 of the model. Regarding magnetic field diffusion, Ohmic and ambipolar diffusion are also negligible in Step 1. The stream's compression throughout the halo could lead to a decrease in its ionisation parameter, resulting in shorter ambipolar diffusion times. However, simulations by \citet{Aung2024} indicate that, with or without self-shielding, the streams maintain a temperature of $\sim 10^4 \, \rm K$, suggesting that the ionisation parameter should not significantly decrease. Ohmic diffusion is also expected to remain insignificant throughout the halo. Eventually, magnetic diffusion is likely to occur inside the galaxy and should be included in Eq.~\ref{eq:dEmgdt2} as a future work. 

Finally, our model does not account for the magnetic field structure in the galaxy. Previous theoretical models \citep[e.g.][]{Henriksen2016,Nixon2018} estimated the resulting structure and polarisation of the magnetic field for a galaxy embedded in a magnetised small halo ($\sim 20 \rm \, kpc$ halo radius). Interestingly, while they do not address the origin of the magnetic field in either the halo or the galaxy, a by-product of their model is an inflow or outflow of magnetic fields. Such an estimation of the galactic magnetic field structure is left for future work.

\section{Conclusions}
We demonstrate that the stretching and tangling of magnetic field lines are effective mechanisms for amplifying the magnetic field within cold streams through their interaction with the CGM. 
Within the framework of the simple assumptions of our model, cold streams emerge as plausible candidates for raising the galactic magnetic field from zero to $>\muG$ for massive high-redshift star-forming galaxies. With reasonable assumptions regarding the mass accretion rate derived from SFR, our models, both with and without outflows, predict a very rapid magnetisation of the galaxy (within $10\, \mathrm{Myr}$) consistent with the $\sim 500\, \muG$ galactic magnetic field observed by \citet{Geach2023}. Such magnetisation timescale could also account for the observed galactic magnetic field at $z=5.6$ \citep{Chen2024}.

\begin{acknowledgement}
N.L. acknowledges support from the European Research Council (ERC) under the European Union’s Horizon 2020 research and innovation program grant agreement No 864361.
% Eva
E.N. acknowledges funding from the Italian Ministry for Universities and Research (MUR) through the "Young Researchers" funding call (Project MSCA 000074). 
Numerical computations were carried out on the Cray XC50 at the Center for Computational Astrophysics, National Astronomical Observatory of Japan, and the {\sc SQUID} at the Cybermedia Center, Osaka University as part of the HPCI system Research Project  (hp230089, hp240141). 
This work is supported in part by the MEXT/JSPS KAKENHI grant numbers  20H00180, 22K21349, 24H00002, and 24H00241 (K.N.). 
K.N. acknowledges the support from the Kavli IPMU, World Premier Research Center Initiative (WPI), UTIAS, the University of Tokyo. 
\end{acknowledgement}

% for the bibliography, at the end
\bibliographystyle{bibtex/aa} % style aa.bst
\bibliography{references.bib} % your references Yourfile.bib

\begin{appendix}
\section{Power spectra}\label{app:power_spectra}
We first present the general evolution of the power spectra of the kinetic energy, the magnetic energy, and the enstrophy, defined as $\omega^2$ with $\omega$ the magnitude of the vorticity. We then quantify the similarities between the functional form of the magnetic energy and the enstrophy power spectra.
\subsection{General analysis}
To confirm that the magnetic field is amplified by stretching and tangling, we compare the energy power spectra\footnote{To save computational time, the power spectra for the 3D simulation are computed at a uniform resolution $(2^{l_{\mathrm{max}}-1})^3$, where $l_{\mathrm{max}}$ the maximum level of refinement in the simulation.} of the magnetic energy $\widehat{e}_{\mathrm{m}}$, kinetic energy $\widehat{e}_{\mathrm{k}}$, and enstrophy $\widehat{\omega^2}$.
\\
\begin{figure*}[!h]
\centering
    \includegraphics[width=2.\columnwidth]{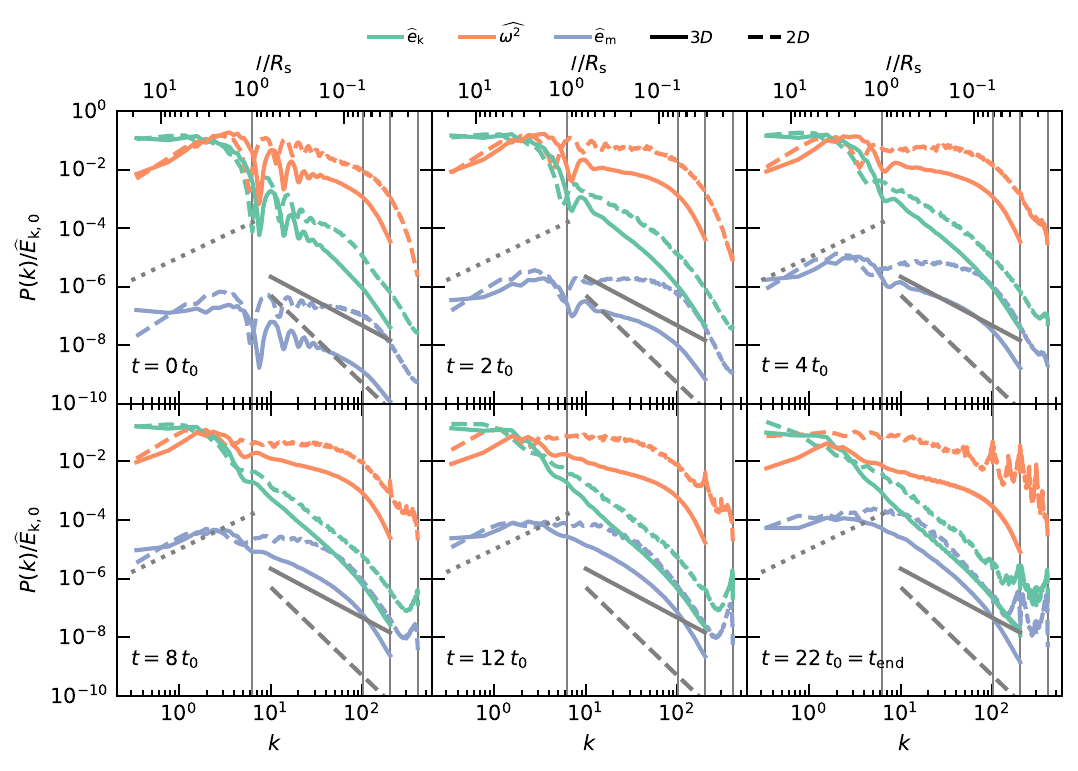}
    \caption{Power spectra of the kinetic energy, the enstrophy, and the magnetic energy normalised by the total initial kinetic energy. Results are shown for both 2D and 3D MHD simulations with $\mathcal{M}=1, \delta=30, n_{\rm{s}}=0.01 \, \rm cm^{-3}$. The spectra are plotted at different times in each panel. From left to right, the vertical black lines refer to length-scales of $R_{\stream}$, $R_{\stream}/16$, $R_{\stream}/32$, and $R_{\stream}/64$, the minimum cell size. For illustration, the power laws $k^{3/2}$, $k^{-5/3}$ and $k^{-3}$ are shown by grey dotted, plain and dashed lines, respectively.}
    \label{fig:Spectrumall}
\end{figure*}
The time evolution of the power spectra for the kinetic energy, magnetic energy, and enstrophy is shown in Fig.~\ref{fig:Spectrumall} for a 2D and a 3D simulation. All spectra are normalised to the initial total kinetic energy $\widehat{E}_{\mathrm{k,0}}= \int \widehat{e}_{\mathrm{k}}\left(t=0\right) \,\mathrm{d}k$. The evolution of the spectra is relatively similar between 2D and 3D, although the energy spectra of the 3D simulation have a steeper slope than in 2D. 
The 2D spectra display artificial peaks at high wavenumber $k$ (green dashed lines for $t \geq 12t_0$), corresponding exactly to the resolution limits of the simulations\footnote{The simulation uses static mesh refinement. Hence, if the stream grows with time, the mixing layer can be shifted to lower resolution regions, leading to the appearance of artificial peaks at 2 and 4 times lower wavelengths than the resolution limit.}. The 3D simulation does not exhibit any artificial peaks. This can be attributed to the following two points: (1) the higher number of cells per wavenumber bin when computing the power spectra in 3D compared to 2D; and (2) the lower resolution of the snapshots used to compute the 3D spectra.

All spectra exhibit an energy injection scale, an inertial range, and a dissipation range. The magnetic energy spectrum shows a typical inverse cascade for scales $\gtrsim R_{\stream}$, analogous to the $k \propto 3/2$ slope predicted by various theoretical models \citep[e.g.,][]{Kazantsev1985,Schekochihin2004}, and as observed in galaxy and cluster simulations \citep[e.g.,][]{Xu2009,Vazza2018,Pakmor2024}. 
A near equipartition \footnote{Equipartition is not strictly observed in Figure~\ref{fig:Spectrumall}. In the simulation, the magnetic field is amplified within the mixing layer, resulting in equipartition only in that region. Since the power spectra are computed over the entire computational domain rather than solely within the mixing layer, this introduces excess kinetic energy into the spectra.}
is first achieved at high wavenumbers, as the magnetic field amplification occurs initially on small scales.In agreement with previous numerical studies on MHD turbulence \citep[e.g.,][]{Ryu2000,Cho2009,Salvesen2014}, the equipartition extends to larger scales over time, resulting in similar slopes for the magnetic and kinetic energy where equipartition is established. The resulting slopes of the kinetic and magnetic energy at high wavenumbers lie between $k^{-5/3}$ and $k^{-3}$.

A detailed characterisation of the slope and its origin through a spectral energy transfer function analysis is beyond the scope of this current work. However, to support that stretching and tangling drive the magnetic field amplification, we present a phenomenological argument: at early times, the magnetic energy spectrum is proportional to the enstrophy spectrum. Initially, the magnetic field is uniform and lacks a defined spectrum, but it immediately adopts the shape of the enstrophy spectrum, corresponding to the characteristic timescale of magnetic field amplification. The magnetic energy spectrum closely follows the shape of the enstrophy spectrum for length scales $\gtrsim 0.1 R_{\stream}$, which also corresponds to the length scales of the mixing layer, where tangling and stretching of the field lines occur. We verify this similarity by quantifying the difference between the magnetic energy and enstrophy spectra in the next section.

\subsection{Quantification of the power spectrum difference}
\begin{figure*}
    \includegraphics[width=2.0\columnwidth]{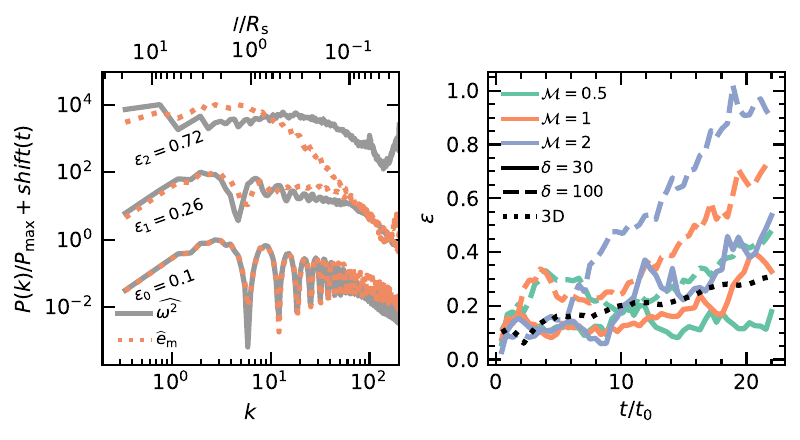}
    \caption{{\it Left:} Illustrative plot of the enstrophy and magnetic energy power spectra normalised by their maximum values at different time $(t_{\rm{0}},t_{\rm{1}},t_{\rm{2}})\sim (0.02,0.1,1) t_{\rm{end}}$. For clarity, the spectra are shifted on the y-axis, and their corresponding deviation $\varepsilon$ is shown. The results shown here are for a 2D MHD+TC simulation with $\mathcal{M}=1, \delta=100, n_{\rm{s}}=0.01 \, \rm m^{-3}$. {\it Right:} Time profile of the difference $\varepsilon$ for different Mach number $\mathcal{M}$, and density ratio $\delta=100$. The results shown here are for 2D MHD+TC simulations with $n_{\rm{s}}=0.01 \, \rm m^{-3}$, and a 3D MHD simulation (black dotted line) with $\mathcal{M}=1, \delta=30, n_{\rm{s}}=0.01 \, \rm m^{-3}$.}
    \label{fig:error}
\end{figure*}

To quantify the difference between the enstrophy and magnetic energy spectrum, both spectra are normalised by their maximum value. From the normalised spectrum $E_{\rm{m,norm}}(k)$ and $\omega^2_{\rm{m,norm}}(k)$, we compute the difference of their logarithmic value,
\begin{equation}
    \varepsilon = \frac{1}{N}\sum_{\rm{N}}\left( \log_{\rm{10}}\left(\frac{E_{\rm{m,norm}}(k)}{\omega^2_{\rm{m,norm}}(k)} \right) \right),
\end{equation}
where $N$ is the number of wavenumber on which the average is done. As we are mainly interested in the short wavenumber, the average is performed for $k<100$. 
Fig.~\ref{fig:error} shows the normalised spectrum at the wavenumbers of interest from one simulation at different times and the time profile of the deviation $\varepsilon$ for different simulations. 
From the normalised spectrum in Fig.~\ref{fig:error}, it appears that the magnetic energy spectrum is nearly identical to the enstrophy one at the early stage of the simulation. The shapes of the spectra slightly deviates from one another towards the end of the simulation. The deviation $\varepsilon$ is shown for different times where it appears that $\varepsilon\sim 0.3$ can still reflect a relatively good agreement between the normalised spectra. 
In the right panel of Fig.~\ref{fig:error}, all simulations exhibit a deviation $\varepsilon \lesssim 0.3$ for $t\lesssim 7t_{\rm{0}}$, which correlates with the growth of the magnetic field being the strongest also for the early stage of the simulation. 
This small value of $\varepsilon$ suggests
%gives phenomenological support 
 that the growth of the magnetic field is driven by tangling and stretching of the field lines at length-scales of size $l\gtrsim 0.1R_{\stream}$.

\section{Evolution of the stream in 2D and 3D}\label{app:comp_2D}
\begin{figure*}
\centering
    \includegraphics[width=2.\columnwidth]{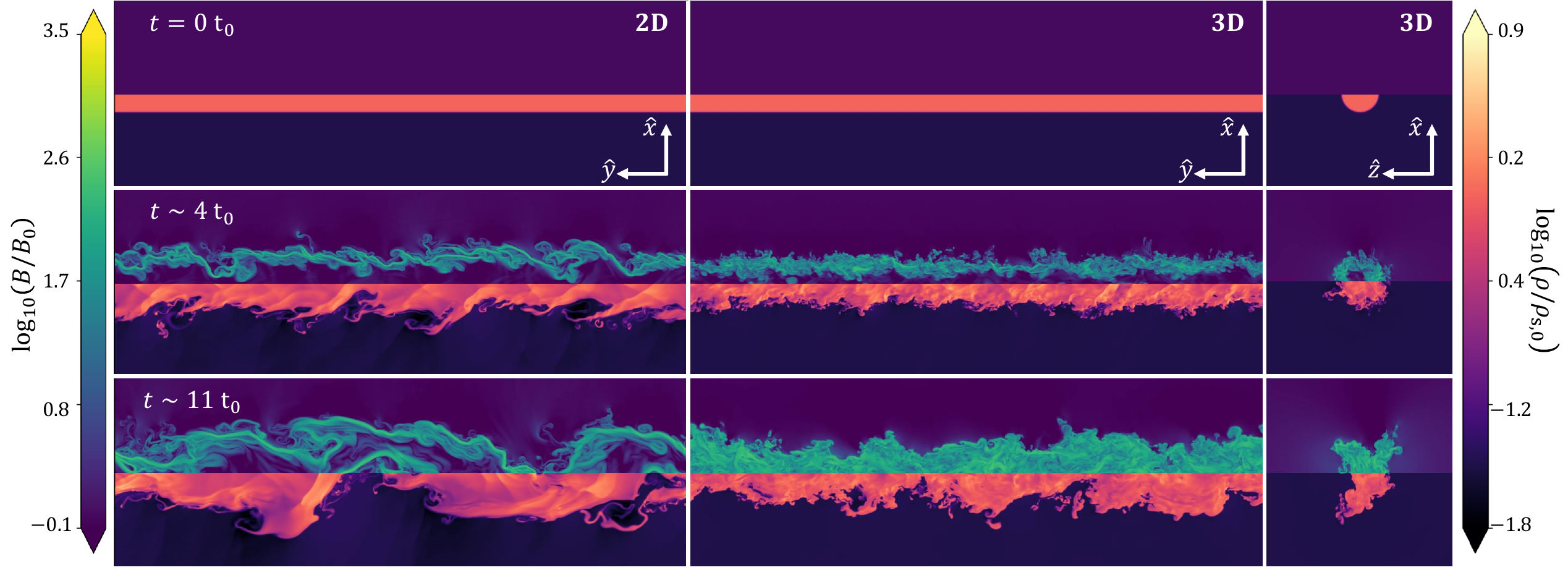}
    \caption{Comparison of magnetic field (top) and density (bottom) in 2D and 3D MHD simulations with $\mathcal{M}=1, \delta=30, n_{\rm{s}}=0.01 \, \rm cm^{-3}$ at the initial time, at $t\sim 4\, t_0$, and at half simulation time $t=11\, t_0$. Each panel shows the $32\,\kpc$ full stream axis length in the horizontal direction and a zoomed region of $10\,\kpc$ in the vertical direction. The 3D maps represent slices of the density and magnetic field and not projected quantities.}
    \label{fig:stream_evo}
\end{figure*}

The mechanisms leading to the magnetic field growth in 2D and 3D simulations are the same. However, slight differences appear in the efficiency of the mixing.
Figure~\ref{fig:stream_evo} shows the time evolution of the stream density and magnetic field strength slices through the stream midplane for both 2D and 3D simulations.
Initially, the magnetic field is amplified at the CGM/stream interface where the velocity shear and the KHI are most pronounced. Consequently, at $t=4 \mathrm{t_0}\sim 0.2 t_{\mathrm{v}}$, in both 2D and 3D simulations, the magnetic field near the centre of the stream remains at its initial value, while it undergoes significant amplification at the interface, reaching up to $\gtrsim 200$ times its initial strength.
By $t=11\,\mathrm{t_0}\sim 0.6 t_{\mathrm{v}}$, the magnetisation of the stream differs between 2D and 3D. In 3D, the mixing is more efficient due to the KHI \citep[see][]{Mandelker2019}, leading to a uniform magnetic field distribution in the stream, while in 2D, the magnetic field near the stream's centre remains unchanged.

\section{Derivation of the model with a simple outflow term}\label{app:deriv_out}
We hereby derive the model assuming that the galaxy experiences a constant outflow from supernovae feedback.

Starting from Eq.~\ref{eq:emg_gen2}, we focus on the inflow and outflow components $E_{\mathrm{m,in}}$ and $E_{\mathrm{m,out}}$, leaving all other terms equal to zero.
We can then rewrite $\dot{E}_{\mathrm{m,g}}$ as
\begin{equation}
    \dfrac{\mathrm{d} E_{\mathrm{m,g}}}{\mathrm{d} t} = \dfrac{\mathrm{d} E_{\mathrm{m,in}}}{\mathrm{d} t} - \dfrac{\mathrm{d} E_{\mathrm{m,out}}}{\mathrm{d} t} = \dfrac{\mathrm{d} E_{\mathrm{m,in}}}{\mathrm{d} M_{\mathrm{in}}}\dfrac{\mathrm{d} M_{\mathrm{in}}}{\mathrm{d} t} - \dfrac{\mathrm{d} E_{\mathrm{m,out}}}{\mathrm{d} M_{\mathrm{out}}}\dfrac{\mathrm{d} M_{\mathrm{out}}}{\mathrm{d} t}.
\end{equation}
$E_{\mathrm{m,in}}$ is still given by Eq.~\ref{eq:emin}, and $E_{\mathrm{m,out}}$ yields,
\begin{equation}\label{eq:emout}
    E_{\mathrm{m,out}}= \frac{B_{\mathrm{g}}^2}{2}  V_{\mathrm{out}} = \frac{E_{\mathrm{m,g}}}{V_{\mathrm{g}}}  \frac{M_{\mathrm{out}}}{\rho_{\mathrm{out}}},
\end{equation}
where $V_{\mathrm{out}}$, $M_{\mathrm{out}}$ and $\rho_{\mathrm{out}}$ are the volume, mass, and density of the gas outflow leaving the galaxy. $V_{\mathrm{g}}$ is the galactic volume such that $E_{\mathrm{m,g}}=0.5V_{\mathrm{g}}B_{\mathrm{g}}^2$. 
We end up with a general formulation as,
\begin{equation}\label{eq:dEmgdtout}
    \dfrac{\mathrm{d} E_{\rm{m,g}}}{\mathrm{d} t} = \frac{B_{\mathrm{in}}^2}{2 \rho_{\mathrm{in}}} \dfrac{\mathrm{d} M_{\mathrm{in}}}{\mathrm{d} t} - \frac{E_{\mathrm{m,g}}}{\rho_{\mathrm{out}}V_{\mathrm{g}}}\dfrac{\mathrm{d} M_{\mathrm{out}}}{\mathrm{d} t} = C_{\mathrm{in}} - \frac{E_{\mathrm{m,g}}}{\tau_{\mathrm{out}}},
\end{equation}
with, $\tau_{\mathrm{out}}$ representing the magnetic field expulsion timescale.
Assuming that $\tau_{\mathrm{out}}$ and $C_{\mathrm{in}}$ are invariant in time, the solution of the first order differential Eq.~\ref{eq:dEmgdtout} is,
\begin{equation}\label{eq:emg_genout}
    E_{\mathrm{m,g}}(t) = E_{\mathrm{m,g,0}}\exp{\left(-\frac{t}{\tau_{\mathrm{out}}}\right)} + \tau_{\mathrm{out}}C_{\mathrm{in}}\left[1-\exp{\left(-\frac{t}{\tau_{\mathrm{out}}}\right)}\right].
\end{equation}
For $C_{\mathrm{in}}$, Eq.~\ref{eq:cin} still holds.
For $\tau_{\mathrm{out}}$, we define the outflow density as the average density of the gas flowing out of the galaxy disc surface $\pi R_{\mathrm{g}}^2$ at a velocity $v_{\mathrm{out}}$,
\begin{equation}\label{eq:dout}
    \rho_{\mathrm{out}} =  \frac{1}{\pi R_{\mathrm{g}}^2 v_{\mathrm{out}}}\dfrac{\mathrm{d} M_{\mathrm{out}}}{\mathrm{d} t} ,
\end{equation}
which leads to,
\begin{equation}\label{eq:cout}
    \tau_{\mathrm{out}}= \frac{2 d_{\mathrm{g}}}{v_{\mathrm{out}}},
\end{equation}
where $d_{\mathrm{g}}$ is the galaxy's disc height, and we fix $v_{\mathrm{out}} = 300 \, \mathrm{km\, s^{-1} }$ based on the isolated galaxy simulations of \citet[][figure 18]{Oku2022}.

Finally, the mean magnetic field in the galactic volume $V_{\rm{g}}$ can be recovered as
\begin{equation}\label{eq:Bgout}
    B_{\rm{g}}(t) = \left(2\frac{ E_{\mathrm{m,g}}(t)}{V_{\rm{g}}} \right)^{1/2}.
\end{equation}
From Eq.~\ref{eq:emg_genout}, the asymptotic value for $E_{\mathrm{m,g}}$ is $\tau_{\mathrm{out}}C_{\mathrm{in}}$ which represents the amount of energy brought by inflow during a timescale $t=\tau_{\mathrm{out}}$. The time for $E_{\mathrm{m,g}}$ to reach a value of $0.9 \tau_{\mathrm{out}}C_{\mathrm{in}}$ is given by $t\sim 2.3\tau_{\mathrm{out}}$.
\end{appendix}

\end{document}